\documentclass[12pt,preprint]{aastex}

\newcommand{\msun}{\mbox{M$_{\odot}$}}
\newcommand{\zsun}{\mbox{Z$_{\odot}$}}

\begin{document}

\title{Three Epochs of Star Formation in the High Redshift Universe}
\author{Jonathan Mackey, Volker Bromm, and Lars Hernquist}
\affil{Harvard-Smithsonian Center for Astrophysics, 60 Garden Street,
Cambridge, MA 02138; jmackey@cfa.harvard.edu, vbromm@cfa.harvard.edu,
lars@cfa.harvard.edu}

\begin{abstract}

We investigate the impact of an early population of massive stars on
their surroundings.  Dissociation of molecular hydrogen by strong UV
emission from such stars is expected to produce a global transition in
the cooling mechanism of minihalos at a redshift of approximately 30,
strongly inhibiting star formation until more massive halos can
collapse.  Furthermore, chemical enrichment from Pop~III supernovae
produces a second transition at $z\sim15-20$, when the mean
metallicity of the universe exceeds a critical threshold and Pop~III
star formation gives way to Pop~II.  We show that observations of high
redshift supernovae with the {\it Next Generation Space Telescope}
(NGST) have the potential to trace the cosmic star formation rate out
to $z\ga 20$, provided that Pop~III supernovae are at least as bright
as, and ideally brighter than, type Ia supernovae.  We also propose a
mechanism for the formation of a novel population of extremely low
metallicity stars of intermediate mass at very high redshifts, which
we term Pop~II.5.  In our model shock compression, heating, and
subsequent cooling to high density reduces the fragment mass in
primordial gas to $\sim10\,\msun$, allowing low mass stars to form.
We predict the number density of relic Pop~II.5 stars in the Milky Way
halo today and find that, with certain assumptions, there should be
$\sim 10\ {\rm kpc^{-3}}$ in the solar neighborhood.

\end{abstract}

\section{Introduction}

An important challenge in modern cosmology is to understand how the
cosmic ``dark ages'' ended \citep[for a recent review,
see][]{BarLoe01}.  There is growing theoretical evidence indicating
that the first luminous objects to form in the universe were very
massive stars with typical masses $M_{\ast}\ga 100\,\msun$
\citep*{BroCopLar99,BroCopLar02,AbeBryNor00,AbeBryNor02,NakUme01}.  
These stars formed out of metal-free gas in dark matter (DM) halos of
mass $\sim 10^5 - 10^6 \;\msun$ at redshifts $z \ga 20$
\citep[e.g.,][]{Tegetal97,FulCou00}.  Typically, simulations predict
that the halos hosting this process should contain either one or a few
dense massive clumps of baryonic matter, with the most massive clump
situated nearest to the center of the halo being the one to collapse
first and to presumably form a very massive star (VMS).  On the other
hand, conditions in the universe less than $10^{9}$ yr after the big
bang must have allowed for the formation of low-mass stars, with
typical masses of $\sim 1 \;\msun$, as is implied by the ages of the
oldest globular clusters in the Galactic halo with metallicities of
$Z\sim 10^{-2}\zsun$ \citep[e.g.,][]{AshZep98,BroCla02}.  In this
paper, we investigate the question: {\it How and when did the
transition from the formation of very massive stars at high redshifts
to that of more normal, low-mass stars at later times take place?}

Observations have recently provided hints to the character of star
formation at $z\ga 5$.  The abundance of C~IV in the low-column
density Ly$\alpha$ forest indicates that the intergalactic medium
(IGM) was enriched with heavy elements to a level of $Z\sim
10^{-3.5}\zsun$ already at $z\sim 5$ \citep{Son01}. In addition,
abundance patterns in Ly$\alpha$ systems out to $z \sim 4.6$ are
suggestive of an early, prompt nucleosynthetic inventory of heavy
elements from a generation of very massive stars \citep*{QiaSarWas02}.

Elucidating the cosmic star formation history in the high redshift
universe is crucial for predicting what observatories such as the {\it
Next Generation Space Telescope} (NGST) will observe at $z > 5$ less
than a decade from now. In particular, it is important to ascertain
the rates and properties of high redshift supernovae (SNe) which are
likely to be the brightest beacons heralding the end of the ``dark
ages'' \citep{MirRee97}. 
Similarly, gamma-ray bursts (GRBs) are expected to trace the formation
history of massive stars out to very high redshifts
\citep{Tot97,Wijetal98,BlaNat00}.  In fact, the top-heavy initial mass
function (IMF) predicted for the first stars favors the massive stars
which are the likely source of GRB progenitors.  Constraining how
stars form at high $z$, therefore, is of great relevance for
interpreting the results of the upcoming {\it Swift} satellite
\citep{BroLoe02}.

Here, we examine processes related to star formation in the high
redshift universe using a simple analytical model that synthesizes a
number of recent theoretical and observational results into a coherent
framework.  This idealized model is intended to emphasize the basic
physical mechanisms, and we will complement it with detailed numerical
simulations in future work.

The organization of the paper is as follows. In \S 2 we describe our
high-redshift star formation model.  A discussion of the various
stellar populations that form at high $z$ is given in \S 3, while \S 4
presents observational consequences. Finally, \S 5 contains our
conclusions and the outlook for future work.

\section{Epochs of Star Formation at High Redshift}
\subsection{Cosmic Star Formation History}

In general, VMSs will impart three forms of feedback on the IGM
\citep[e.g.,][]{Ciaetal00}: radiative, chemical, and mechanical.  Owing
to their high effective temperatures, $T_{\rm eff}\sim 10^{5}$K, these
stars produce a large quantity of UV photons, roughly 10 times as many
per stellar baryon as those from a ``normal'' stellar population
characterized by a Salpeter IMF \citep*{TumShu00,BroKudLoe01}.  Upon
its death after only $\sim 3\times 10^{6}$yr, a VMS explodes as a
supernova and enriches the IGM with heavy elements, provided the star
is not more massive than $\sim 260 \;\msun$, in which case it is
predicted to collapse entirely into a massive black hole without any
concomitant metal ejection \citep*{FryWooHeg01}.  We propose that this
radiative and chemical feedback gives rise to two fundamental
transitions in the cosmic star formation history at high $z$.  The
feedback from mechanical energy input into the IGM might lead to a
distinct new population of stars, to be discussed in \S 3.

\begin{figure} 
\plotone{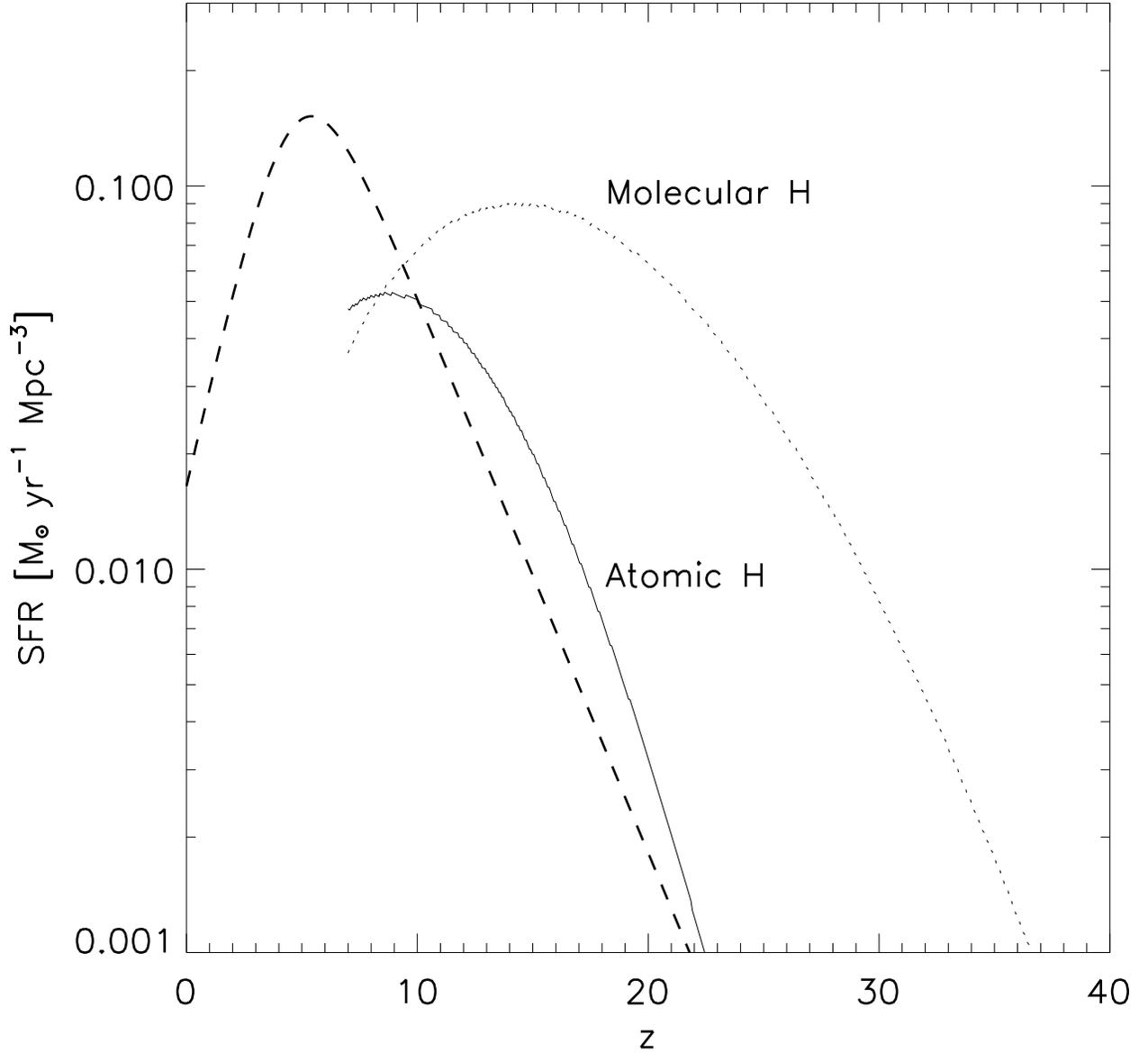} 
\caption{History of comoving
star formation rate (SFR) density
in units of $\;\msun$yr$^{-1}$Mpc$^{-1}$
as a function of redshift.  
{\it Solid line:} Formation rate of VMSs in halos that cool due to atomic
hydrogen.
{\it Dotted line:} VMS formation rate in halos that cool due to molecular
hydrogen.
In calculating these rates, VMS formation was assumed to proceed
in a single-burst mode with an efficiency of
$\eta_\ast=10\%$ \citep[see][]{SanBroKam02}.
{\it Dashed line:} Analytic fit to the star formation history
derived from cosmological simulations that incorporate a multi-phase
model of quiescent star formation \citep[from][]{SprHer02}.
\label{fig1}}
\end{figure}  

To assess the global feedback exerted by VMSs on the IGM, we need to
know their formation rate.  We calculate this rate according to the
``single-burst'' model of VMS formation introduced by
\citet*{SanBroKam02}.  This model adopts the observationally favored
scenario in which the formation of cosmic structure has progressed
hierarchically from small to large scales as described by the cold
dark matter (CDM) model with a cosmological constant.  Specifically,
we assume a $\Lambda$CDM model with density parameters in matter
$\Omega_m=1-\Omega_{\Lambda}=0.3$ and in baryons $\Omega_B=0.045$, a
Hubble parameter of $h=H_0/(100 {\rm km\,s^{-1}\,Mpc^{-1}})=0.7$, and
a scale-invariant power spectrum of density fluctuations with a
normalization $\sigma_8 = 0.9$.  The abundance and merger history of
the DM halos is given by the extended Press-Schechter formalism
\citep{LacCol93}. Within merging halos, stars form in gas that is able
to cool efficiently, via either molecular or atomic hydrogen
lines. These cooling mechanisms become effective at different minimum
temperatures, $T_{\rm crit}\sim 400$ K and $\sim 10^{4}$K for
molecular and atomic hydrogen, respectively \citep[e.g.,][]{BarLoe01}.
These temperature thresholds correspond to redshift-dependent minimum
DM halo masses
\begin{equation}
M_{\rm crit}(z)\simeq 10^{8}\;\msun\left(\frac{\mu}{0.6}\right)^{-3/2}
\left(\frac{T_{\rm crit}}{10^{4}\mbox{K}}\right)^{3/2}
\left(\frac{1+z}{10}\right)^{-3/2}\mbox{\ ,}
\end{equation}
where $\mu$ is the mean molecular weight ($\mu=0.6$ for ionized gas,
and $\mu=1.2$ for neutral gas). Finally, the star formation efficiency
is chosen to be $\eta_{\ast}=0.10$.
The ``single-burst'' model now assumes that VMS formation is accompanied
by strong negative feedback effects on the star-forming gas (see below
for a detailed discussion). In essence, VMSs are only allowed to form
out of gas that has never perviously experienced star formation, in
merging DM halos that cross $M_{\rm crit}(z)$ for the first time.

We only consider mergers in which the two progenitor halos are within
a factor of 100 of each other in mass, although our results are quite
similar (within a factor of 2) if this ratio is as low as 10 or as high
as 300.  We impose this limit because very minor mergers will not set
off a burst of VMS formation, and also because they take a long time.
Dynamical friction cannot cause halos to merge within a Hubble
time if their mass ratio is large.  Any uncertainty in the allowed
mass ratio is significantly smaller than the uncertainty in the
value of $\eta_*$.

Figure~\ref{fig1} shows the resulting star formation histories.  We here
presume that VMS formation will have ceased at latest by $z\sim 7$,
close to the estimated redshift of reionization 
\citep[][and references therein]{Bar02}. The precise redshift where 
VMS formation ends is uncertain, and is determined by the feedback
effects that we will discuss below.  For the sake of comparison, in
Figure~\ref{fig1} we also show an analytical fit to the star formation history
obtained by \citet{SprHer02} from a series of cosmological simulations
that incorporate a self-regulated model of quiescent star formation in
a multi-phase interstellar medium.  This latter history describes the
formation of a normal, predominantly low-mass population of stars (see
\S 3).  Note that the Springel \& Hernquist results were obtained for
a cosmology with a slightly smaller value of $\Omega_B=0.04$ than for
our current analysis.

\subsection{Radiative Feedback and the First Transition
in the Star Formation History}

In a hierarchical model of structure formation, the very first stars
will form in small DM halos where gas cooling has to rely on the
presence of molecular hydrogen. Initially, therefore, star formation
proceeds along the molecular branch in Figure~\ref{fig1}, and is characterized
by a top-heavy IMF.  This latter prediction follows from the
microphysics of H$_{2}$ cooling, leading to characteristic values of
the gas temperature and density, thus imprinting a characteristic mass
scale of a few $100 \,\msun$ \citep{BroCopLar02}.  The very first
stars derive from $3-4\sigma$ peaks and, albeit rare, efficiently
produce Lyman-Werner (LW) photons with energies (11.2-13.6eV) just
below the hydrogen ionization threshold.  The LW photons are therefore
free to propagate through an otherwise neutral IGM, readily destroying
H$_{2}$ in neighboring halos
(\citealt*{HaiReeLoe97,HaiAbeRee00,RicGneShu01}; but see
\citealt*{CiaFerAbe00}).

Very massive stars produce approximately $\sim 10^{4}$ LW photons per
stellar baryon \citep{BroKudLoe01}.
Roughly, one can then estimate the redshift where star formation in
low-mass halos (relying on H$_{2}$ cooling) is no longer possible by
demanding that there be of order one dissociating LW photon for each molecule
in the IGM:
\begin{equation}
10^{4} f_{\rm LW} f_{\rm diss}
\rho_{\ast}(z_{1\rightarrow 2}) \sim f_{{\rm H}_{2}}\rho_{\rm B}\mbox{\ .}
\end{equation}
Here, $\rho_{\rm B}=4.1\times 10^{-31}$g cm$^{-3}$ is the comoving
density in baryons, $f_{\rm LW}$ the escape fraction of LW
photons from the star-forming halo, $f_{\rm diss}$ the fraction of 
absorptions that lead to dissociation, 
$f_{{\rm H}_{2}}$ the molecule fraction
in the IGM, and
\begin{equation}
\rho_{\ast}(z) =
\int_{z}^{\infty}\psi_{\ast}(z')\left|\frac{{\rm d}t}{{\rm d}z'}\right|
{\rm d}z'
\end{equation}
is the comoving density in stars, with $\psi_{\ast}(z)$ being the
comoving star formation rate (SFR), in this case given by the molecular
track in Figure~\ref{fig1}, and the other symbols have their usual meaning 
\citep[e.g.,][]{BroLoe02}.

\begin{figure} 
\plotone{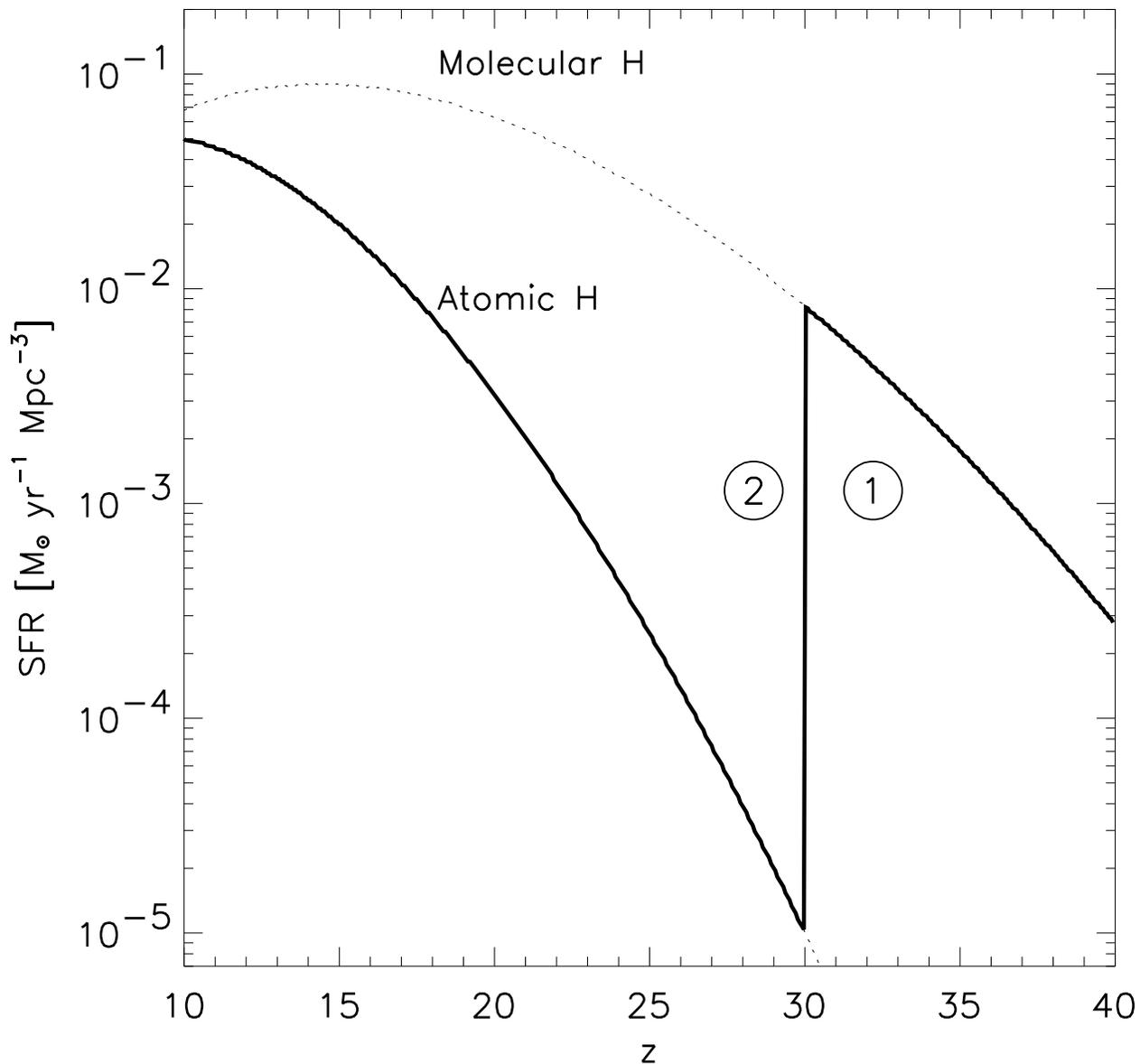}
\caption{The transition from epoch 1 to 2.
Comoving star formation rate (SFR)
in units of $\;\msun$yr$^{-1}$Mpc$^{-3}$
as a function of redshift.  
{\it Dotted lines:} VMS formation rate calculated for cooling due to atomic
and to molecular hydrogen, respectively (from Figure 1).
{\it Solid line:} The actually realized star formation history.
Once LW-photodissociation feedback becomes effective at $z\sim 30$, the SFR
jumps from the molecular to the atomic track. This jump marks the transition
from the first to the second epoch of high-$z$ star formation.
\label{fig2}}
\end{figure} 

In Figure~\ref{fig2}, we evaluate this criterion with $f_{\rm LW}=0.1$,
$f_{\rm diss}=0.1$ (Glover \& Brand 2001) and
$f_{{\rm H}_{2}}=10^{-3}$, finding $z_{1\rightarrow 2}\ga 30$. We
interpret this redshift as the approximate boundary between the first
two epochs of cosmic star formation.  Notice that this is a rather
conservative estimate, as the fraction of gas in molecular form is only
$\sim 10^{-6}-10^{-3}$, and the required number of LW photons might
therefore be significantly smaller than the value derived from
equation (2). It is also worth pointing out that such an early
transition redshift is a direct consequence of the predicted top-heavy
IMF for the first stars, with a factor $\ga 10$ enhancement in the
production of LW photons per stellar baryon compared to the case of a
standard Salpeter IMF \citep{BroKudLoe01}. We have verified that
this transition redshift is not very sensitive to variations within
factors of a few in $f_{\rm LW}$ and $f_{\rm diss}$.

At lower redshifts, star formation can now only proceed in halos that
are massive enough to cool via atomic hydrogen lines (corresponding to
masses $M\ga 10^{8} \;\msun$).  To form stars, the gas has to cool
below the $\sim 10^{4}$K that can be reached by atomic line
cooling. Ultimately, therefore, star formation in these massive halos
still relies on cooling due to H$_{2}$ which is expected to form in
dense regions that are shielded from the LW background
\citep{OhHai02}. The same physical reasoning that leads to the
prediction of a top-heavy IMF in small, $\sim 10^{6}\;\msun$, halos
should then also apply to these more massive halos, regardless of
whether H$_{2}$ is initially present in them or not. The formation of
stars along the atomic track (see Figure~\ref{fig2}) during epoch 2
consequently proceeds with the same top-heavy IMF as in epoch 1.  

The
break in the star formation rate at $z=30$ is artificially sharp due
to the simplicity of our modeling.  In reality, this is likely to be a
smoother transition because the molecular gas will not be dissociated
instantaneously, and will not be completely dissociated in more
massive halos \citep*{MacBryAbe01}.
We do believe, however, that this transition will be rapid.  The mean
free path of LW photons at $z=30$ is $\lambda_{\rm mfp} \sim 10
\;{\rm Mpc}$ (physical), much larger than the typical separation of
$10^6 \;\msun$ halos at this redshift ($\sim 30 \;{\rm kpc}$).
Photons travel this mean free path in $\sim 10^7 \;{\rm yr}$, which is short
compared to the Hubble time.
For these reasons, a roughly uniform background of LW photons
should be set up very quickly.  The number of sources increases
exponentially with time, so the background radiation level will
increase rapidly until it is intense enough to dissociate most of the
molecules in the universe.

The timescale for photodissociation of a molecule is inversely proportional
to the intensity of the radiation field in the Lyman-Werner bands, 
$J_{\rm LW}$, and also depends on how well
the molecule is shielded from the radiation.
\citet{OhHai02} give the relationship as
\begin{equation}
   t_{\rm diss} = 2.1\times 10^{4} J_{21}^{-1} f_{\rm shield}^{-1} \;{\rm yr}
   \ ,
\end{equation}
where $J_{21}= J_{\rm LW}/10^{-21} \;{\rm erg\,cm^{-2}\,s^{-1}\,Hz^{-1}\,sr^{-1}}$,
and $f_{\rm shield}$ takes into account the effects of ${\rm H}_2$ 
self-shielding and is given by 
$f_{\rm shield} = {\rm min}[1, (N_{\rm H_{2}}/10^{14}{\rm cm^{-2}})^{-0.75}]$
(Draine \& Bertoldi 1996).
Thus, the dissociation time at fixed column density depends only on the 
background radiation intensity, approximately given by
\begin{equation}
  J_{\rm LW} \simeq {10^4 h c \over 4\pi m_{\rm H}} \rho_*(z) \exp[-\tau_{\rm IGM}] \ ,
\end{equation}
assuming $10^4$ LW photons are produced per baryon of star formation 
(see above).
The constants in this expression have their usual meaning.
The optical depth of the IGM to LW photons is estimated to be 
$\tau_{\rm IGM} \simeq 2-3$ \citep[e.g.,][]{CiaFerAbe00,RicGneShu01},
reducing the UV flux by about an order of magnitude.
Our formula for $J_{\rm LW}$ neglects the redshifting of photons out of the 
LW bands, but photons travel a mean free path in a small fraction of
the Hubble time and so should be absorbed long before redshift effects 
become important.

\begin{figure} 
\plotone{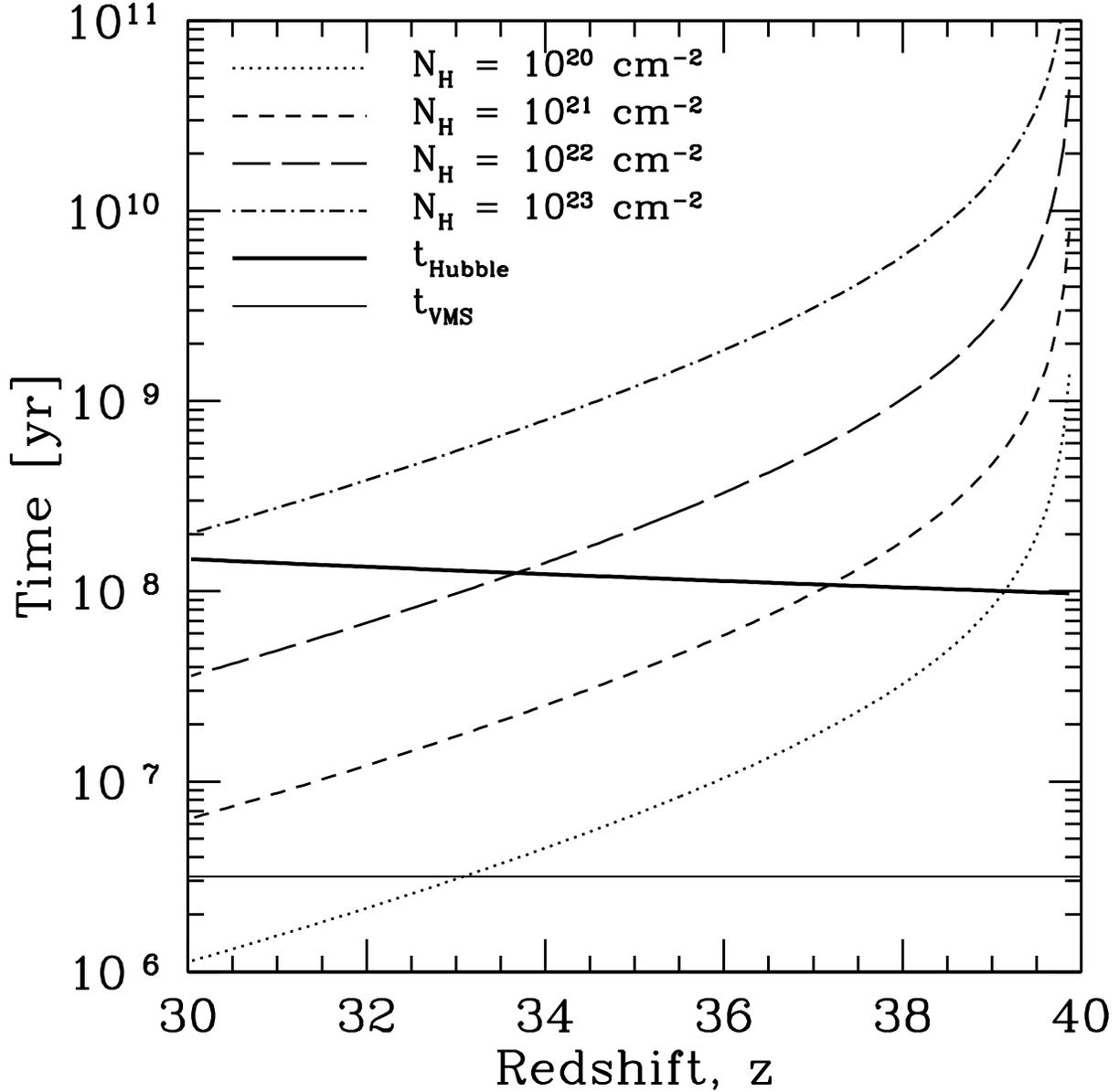}
\caption{The photodissociation time for molecular hydrogen as a function of 
redshift.  The heavy solid line is the Hubble time, the light solid line 
is the lifetime of a VMS ($\sim 3$Myr), and the four broken lines are the
dissociation times in halos with different neutral hydrogen column densities.
Dissociation times decrease towards lower redshifts because the soft-UV
background radiation is increasing rapidly in intensity.
\label{fig:tdiss}}
\end{figure} 

In Figure~\ref{fig:tdiss} we show the dissociation time compared to
the Hubble time as a function of redshift for different neutral hydrogen
column densities.  We also show the lifetime of a VMS for comparison.
We have calculated the curves in Figure~\ref{fig:tdiss} for a
molecular fraction of $10^{-3}$, and with $\tau_{\rm IGM} = 2$.  A
$10^6\;\msun$ halo at $z=30$ would have a hydrogen column density of 
$\sim 4\times10^{20} \;{\rm cm^{-2}}$ if it were a uniform sphere, and
about 10 times larger if it had an isothermal density distribution
\citep{GloBra01}.  To relate this to the rarity of the peaks, we
again consider $z=30$.  A 3-sigma peak has a mass of only 
$4\times 10^3 \;\msun$, and so does not collapse. The
extremely rare peaks in the density field are thus responsible for
dissociating the hydrogen molecules. The masses of halos corresponding to 
4-, 5-, and 6-sigma peaks are
$\sim 5\times 10^5,\ 1\times10^7,\ {\rm and}\ 1\times 10^8 \;\msun$,
respectively.  For a truncated isothermal sphere, this corresponds to
column densities towards the center of the halos of 
$N_{\rm H} \sim 6\times10^{21},\ 2\times10^{22},\ {\rm and}\
4\times10^{22} \;{\rm cm}^{-2}$.  Comparing these column densities 
with the respective curves in
Figure~\ref{fig:tdiss}, we see that the dissociation timescales are
always less than the Hubble time for even the rarest objects by
$z=30$.  Thus we tentatively conclude that this transition is indeed likely to
be rapid.

\subsection{Chemical Feedback and the Second Transition
in the Star Formation History}

Here, we make the simplifying assumption that {\it all} VMSs have
masses in the range $140 \la M \la 260 \;\msun$, and that they
consequently explode as pair-instability supernovae (PISNe) which
completely disrupt the star without leaving behind a compact remnant
(\citealt{FryWooHeg01}; but see \citealt{UmeNom02}). 
PISNe are predicted to have substantial metal
yields, of order $y=M_{Z}/M_{\ast} \sim 0.50$ \citep{HegWoo02}.  
The heavy elements that are produced in the
SN explosions will enrich the host halo, and a fraction, $f_{\rm
mix}$, of them will escape into the general IGM as a result of
SN-driven outflows \citep*[e.g.,][]{MadFerRee01,MorFerMad02,ThaScaDav02}.  
While in \S 3 we predict that some low
mass star formation will occur simultaneously with VMS formation, 
these stars
will explode as type II supernovae which have a much lower metal
yield ($\sim1\msun$ per supernova).  This means that their
contribution to the metallicity of the universe is small as long as
there is comparable star formation in VMSs, and we thus ignore
any contribution from low mass stars in the following calculation.

\begin{figure} 
\plotone{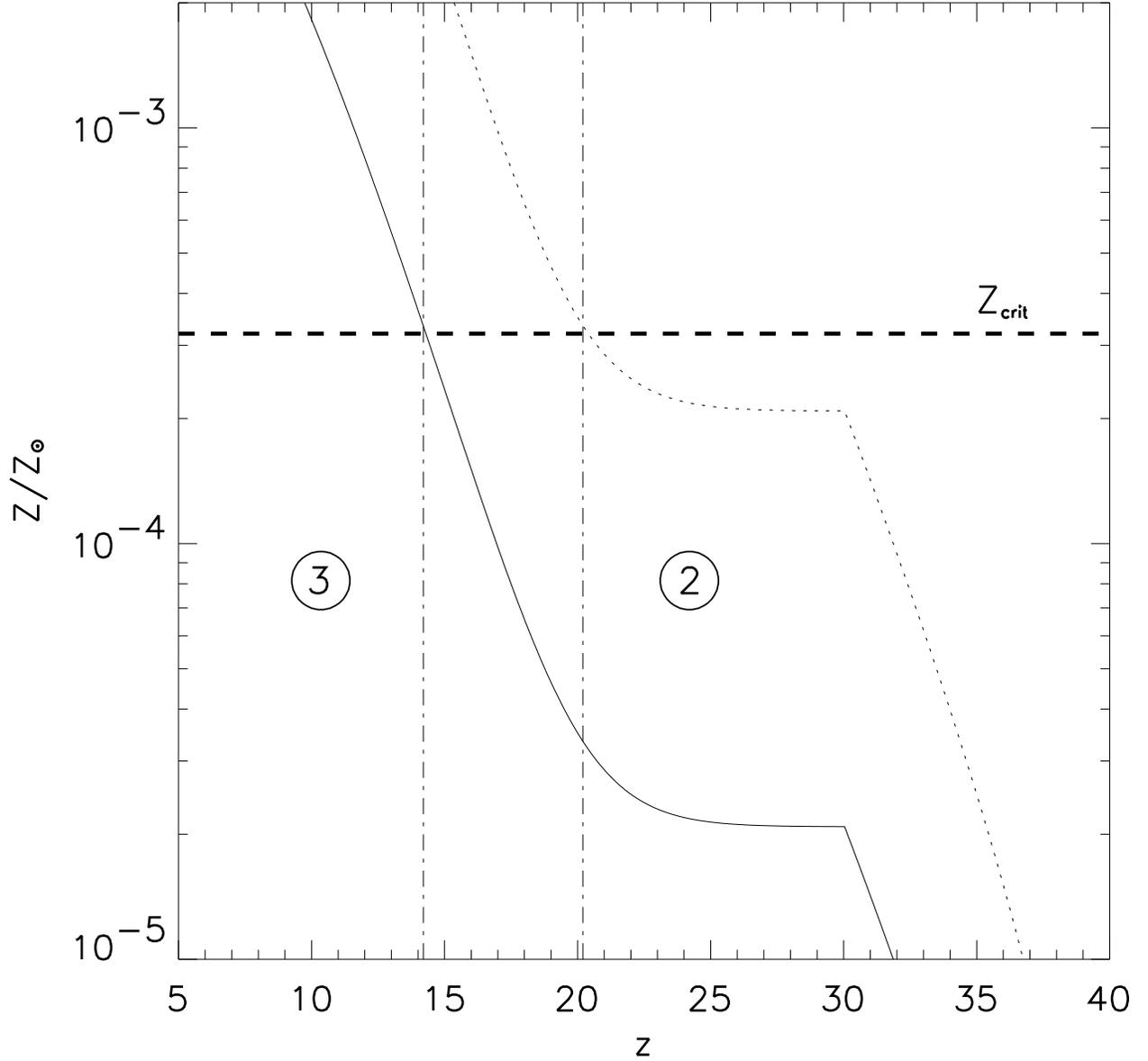}
\caption{The transition from epoch 2 to 3.
IGM metallicity
in units of $\zsun$
as a function of redshift.
The metallicity is calculated with the star formation history in Fig.~1,
assuming a heavy element yield of $50\%$, and two different mixing
efficiencies.
{\it Solid line:} Metal enrichment for $f_{\rm mix}=0.05$.
{\it Dotted line:} Metal enrichment for $f_{\rm mix}=0.5$.
{\it Horizontal dashed line:} The critical metallicity above which massive stars
are no longer able to form.
{\it Vertical dot-dashed lines:} The redshift at which the IGM metal
enrichment reaches $Z_{\rm crit}$, marking the transition from the second
to the third epoch of star formation at high~$z$.
These two lines bracket the plausible range, $15 \la z \la 20$, where the
transition is likely to occur.
\label{fig:metal}}
\end{figure}

With the star formation history described in Figure~\ref{fig2}, we now
calculate the resulting metal enrichment of the IGM
\begin{equation}
Z_{\rm IGM}(z) = y f_{\rm mix} \frac{\rho_{\ast}(z)}{\rho_{\rm B}}
\mbox{\ ,}
\end{equation}
where $\rho_{\ast}(z)$ is the cumulative density in stars as given in
equation (3).  The crucial uncertainty is the fraction, $f_{\rm mix}$,
of metals that are able to escape the star forming system, and to get
mixed into the general IGM. For simplicity, we assume that all the
metals escaping a star forming halo are mixed uniformly into the
IGM. To bracket the likely cases, we calculate $Z_{\rm IGM}(z)$ for
two values of the mixing efficiency: $f_{\rm mix}=0.05$ and 0.5.  It
appears likely that the presence of metals is the key determinant in
ending the epoch of high-mass star formation
\citep{Omu00,Broetal01}. These authors find that very massive stars
can no longer form once the gas is enriched to a level in excess of
$Z_{\rm crit}\simeq 10^{-3.5} \zsun$. The criterion for the second
transition in the cosmic star formation history is then
\begin{equation}
Z_{\rm IGM}(z_{2\rightarrow 3}) \sim Z_{\rm crit}\mbox{\ .}
\end{equation}
In Figure~\ref{fig:metal}, we show that this condition results in the approximate
range, $15 \la z_{2\rightarrow 3} \la 20$, with the limits
corresponding to the two assumptions for the mixing efficiency.  In a
somewhat different context, \citet{Schetal02} have carried out a
similar calculation, finding a transition redshift, $z\ga 20$, that is
broadly consistent with our estimated range.

In reality, not all locations in the universe will have a metallicity
close to the mean value calculated in equation (6). In fact,
the gas in a star forming halo is likely to be rapidly enriched
to a level $\sim Z_{\rm crit}$ locally (see \S 3.3), in contrast
to the rather slow enrichment of the general IGM. A substantial
fraction of the Pop~III metal production, however, is expected to
escape from the shallow DM potential wells at high redshifts (e.g.,
Madau et al. 2001; Mori et al. 2002). In addition, the enriched gas
has to travel much shorter distances between neigboring halos
at these early times, and it might therefore have been easier
to establish a uniform metal distribution in the IGM.
The mean IGM metallicity in equation (6) would then indeed be a 
good indicator of when the universe as a whole will undergo the
transition from predominantly Pop~III to predominantly Pop~II star
formation. Clearly, a more realistic understanding of how the metals
are mixed, and to what degree of homogeneity, can only be gained from
numerical simulations.

In addition to the metal enrichment due to supernovae, the IGM
could also be enriched due to winds from the first stars.
The importance of such winds, however, is predicted to be small
in the absence of metals and dust grains in the stellar atmosphere
(e.g., Kudritzki 2002).

\subsection{Implications}

The sequence of star-forming modes proposed in this section is
expected to apply equally to an individual halo (going through
different `stages' of star formation), and in a statistical sense to
the universe as a whole (going through different `epochs'). It will be
very interesting to determine the degree of synchronization in the
star-formation modes between spatially disjoint regions of the
universe. Is it possible that stars in a given location form in a
different mode from that in a neighboring region? Or is the VMS
feedback sudden enough to coordinate the star formation activity in a
global fashion?  We emphasize that answers to these important questions
must await the results from full numerical simulations, and cannot be
obtained from the simple analytical model presented in this paper.

By examining Figure~\ref{fig:metal}, it appears that the IGM could reach a
level of metal enrichment from VMS formation which is already close
to that inferred for the low-column density Ly$\alpha$ forest at
$z\sim 5$ \citep{Son01}. Within the context of our model,
the IGM would have been enriched with a `bedrock' of metals,
corresponding to $Z_{\rm IGM}\sim Z_{\rm crit}$, already at
$z_{2\rightarrow 3}\ga 15$. Again, a more realistic assessment of this
hypothesis requires sophisticated numerical simulations.

\section{Stellar Populations at High z}

\subsection{Three Basic Populations}

As a consequence of our picture for high redshift star formation, we
predict three distinct stellar populations at $z\ga 5$ (see Table
1). VMSs form out of gas with $Z<Z_{\rm crit}$ and constitute the
`classic' Population~III.  Low-mass stars, with typical masses of
$\sim 1 \;\msun$, that form out of already enriched gas with $Z>
Z_{\rm crit}$ make up Population~II.  These stars are actually
observed, e.g., in the halo of our Galaxy. Finally, we propose a
mechanism for producing a novel Population~II.5: intermediate-mass
stars, forming out of gas with $Z<Z_{\rm crit}$, that owe their
existence to the mechanical, SN-driven feedback from VMSs.


\begin{deluxetable}{lllcc}
\footnotesize
\tablewidth{13.cm}
\tablecaption{Stellar Populations at High Redshifts \label{tab1}}
\tablecolumns{8}
\tablehead{
\colhead{} &
\colhead{$M_{\ast} / \msun$} &
\colhead{$Z / \zsun$} &  
\colhead{QW SNe$^{\rm a}$} &
\colhead{Epoch}  
 }
\startdata
Pop~III...... & $\ga 100$ & $< Z_{\rm crit}$ & VMS & 1,2 \\
Pop~II.5..... & $\ga 10$ & $< Z_{\rm crit}$ & SN~II({\it L}) & 2 \\
Pop~II....... & $\ga 1$ & $> Z_{\rm crit}$ & SN~II({\it H}),({\it L}) & 2,3 \\
\enddata
\tablenotetext{a}{Tentative identification with the different kinds
of SNe proposed by \citet{QiaWas02}.}
\end{deluxetable}

\subsection{The Formation of Population~II.5 Stars} \label{sec:pop2.5}

We assume that the immediate progenitor of a star is a centrally
concentrated, self-gravitating gaseous clump with a mass close to the
Bonnor-Ebert value \citep[e.g.,][]{Pal02}
\begin{equation}
M_{\rm BE} \simeq 700 \;\msun \left({T \over 200{\rm K}}\right)^{3/2}
        \left({n \over 10^{4} {\rm cm}^{-3}}\right)^{-1/2}.
\end{equation}
Here, the temperature and density are normalized to the characteristic
values that follow from the microphysics of H$_{2}$ cooling
\citep[see][]{BroCopLar02}.  Depending on the details of how the gas
is accreted onto the nascent hydrostatic core in the center of the
gravitationally unstable clump \citep[e.g.,][]{OmuPal01,RHFC02}, the
resulting stellar mass is expected to be somewhat smaller. We take
this uncertainty into account by expressing the final stellar mass as
$M_{\ast}\simeq \alpha M_{\rm BE}$, and choose the efficiency to be
$\alpha \sim 0.3$. This value is close to that inferred for the
formation of stars in the present-day universe \citep{McKTan02}.  For
Pop~III stars we then have $M_{\ast}\ga 200 \;\msun$, in agreement
with the results from numerical simulations.  Notice that the
effective star formation efficiency introduced in \S~2.1 is given by
$\eta_{\ast}\simeq f_{\rm cool}\alpha$, where $f_{\rm cool}$ is the
fraction of the gas that is incorporated into the cold, dense clumps.

As pointed out in \S~2.3, the primordial gas cannot cool below $\sim
200$ K as long as $Z < Z_{\rm crit}$.  A decrease in the stellar mass,
therefore, has to rely on achieving higher densities, beyond the value
of $\sim 10^{4}$cm$^{-3}$ that characterizes Pop~III star formation.
We propose that such a boost in density can naturally occur in
the cooled, dense shell behind a SN-driven shock wave, in the
following way.

An exploding VMS drives a blast-wave into the surrounding medium that
shock-heats the gas to a post-shock temperature approximately given by
$T_{\rm ps}\simeq (m_{\rm H}/k_{\rm B}) u_{\rm sh}^{2}$, where $u_{\rm
sh}$ is the (time-dependent) shock speed. Initially, the shock will
evolve adiabatically with only a modest increase in density, $n_{\rm
ps}/n_{0}\leq (\gamma + 1)/(\gamma -1)=4$ for $\gamma=5/3$.  Here,
$n_{0}$ is the density of the undisturbed, preshock gas.  Eventually,
however, the shocked gas will be able to cool and, evolving roughly
isobarically \citep[e.g.,][]{ShaKan87,YamNis98}, reach much higher
compression factors.

\begin{figure} 
\plotone{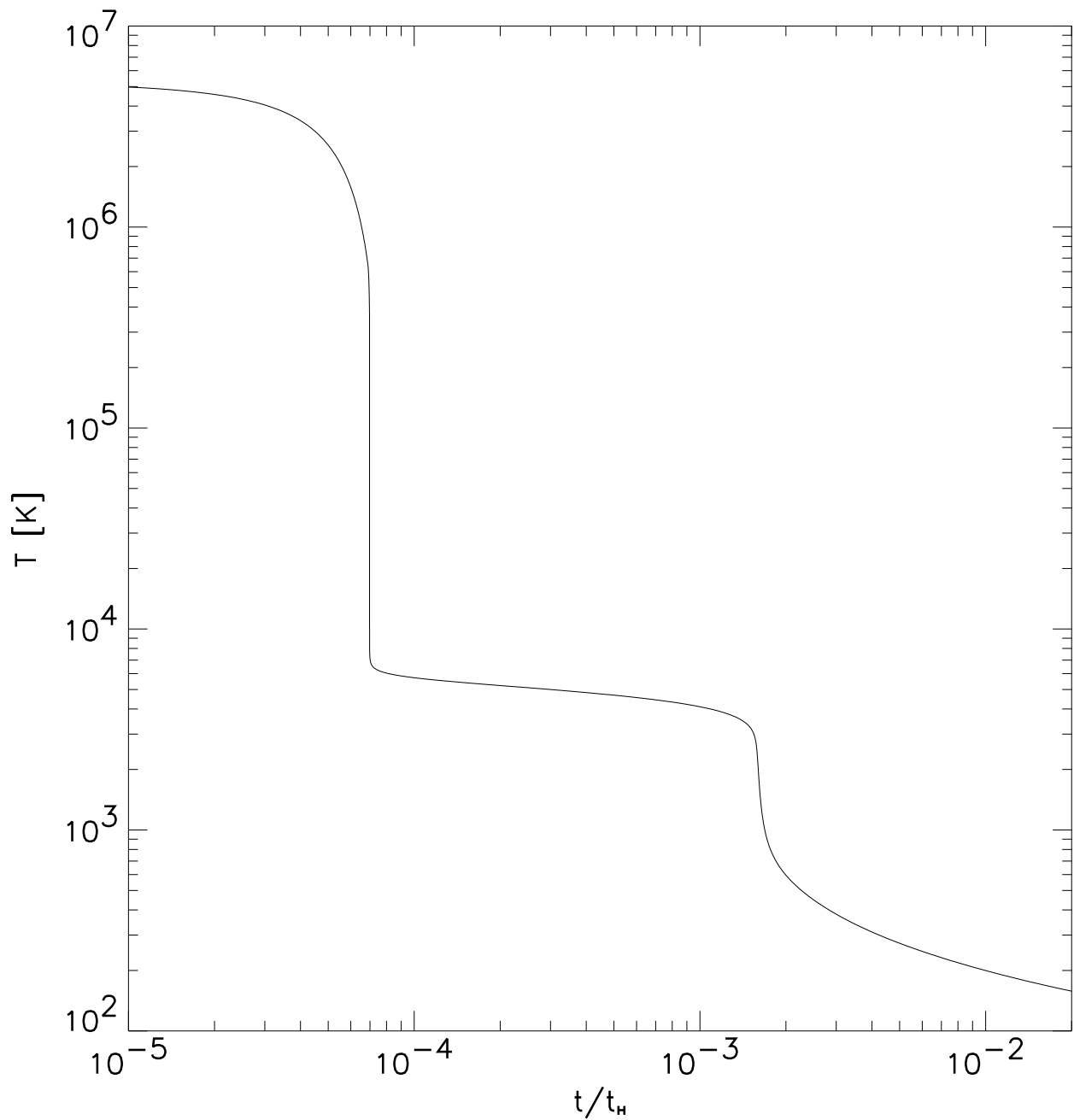}
\caption{Thermal evolution in shocked gas.
Gas temperature
vs. time (in units of the Hubble time at $z\simeq 20$).
Here, the shock has a velocity of $u_{\rm sh}= 200$ km s$^{-1}$, and
propagates into a medium with density $n_{0}=10^{2}$cm$^{-3}$.
The initial, very sudden decrease in temperature is due to the onset
of atomic (He and H) line cooling. The subsequent drop is brought
about by molecular (H$_{2}$) cooling. It is evident that the
shocked gas is able to cool to $\sim 200$ K within a small fraction
of the Hubble time at $z=20$.
\label{fig:cool}}
\end{figure}

To assess whether the post-shock gas can cool sufficiently, we have
implemented a simple one-zone model, solving for the thermal and
chemical evolution in a homogeneous parcel of gas. The chemical
reaction network and the cooling processes pertaining to primordial
gas are described in \citet{BroCopLar02}. Under the assumption of
isobaricity, the density evolves according to $n\simeq n_{0}T_{\rm
ps}/T$. Assuming that the post-shock gas is initially fully ionized,
the only free model parameters are $n_{0}$ and $u_{\rm sh}$.  In
Figure~\ref{fig:cool}, we show the resulting temperature evolution for
$n_{0}=10^{2}$cm$^{-3}$, characteristic of the gas in the vicinity of
a VMS \citep[see][]{BroCopLar02}, and $u_{\rm sh}= 200$ km s$^{-1}$,
corresponding to $T_{\rm ps}\simeq 5\times 10^{6}$K. In this case, the
gas cools to $T_{f}\sim 200$ K within $t\leq 10^{-2} t_H (z=20) \sim
10^{6}$yr, which is shorter than the lifetime of a VMS. The overall
compression in the case of Figure~\ref{fig:cool} is then $n_{f}\simeq n_{0}T_{\rm
ps}/T_{f}\simeq 10^{4}n_{0}$.  In general, the characteristic mass of
a Pop~II.5 star can be written as
\begin{equation}
M_{\ast}({\rm II.5}) \simeq 10 \;\msun 
        \left({n_0 \over 10^{2}{\rm cm}^{-3}}\right)^{-1/2}
        \left({u_{\rm sh} \over 200 {\rm km s}^{-1}}\right)^{-1}.
\end{equation}
This expression implicitly assumes that the post-shock gas is able to
cool to $T_{f}\sim 200$ K in a time that is short compared to the
Hubble time.  This requirement constrains the shock speed to be
$u_{\rm sh}\la 300 {\rm km s}^{-1}$.  The characteristic mass of a
Pop~II.5 star is consequently reduced by at least an order of
magnitude compared to the Pop~III case.

Within our model, the formation of Pop~II.5 stars is directly linked
to the death of a Pop~III VMS. An important quantity then is the
efficiency, $\eta$, of the formation process which we define as the
mass ratio of intermediate mass (Pop~II.5) stars to very massive
(Pop~III) stars
\begin{equation}
    \eta \equiv {M_{\rm II.5} \over M_{\rm III}}
    \sim {\epsilon \ \Delta M_{\rm shell} \over M_{\ast}({\rm III})}
     \;,
\label{eqn:efficiency}
\end{equation}
where $\Delta M_{\rm shell}$ is the mass of the cooled, dense shell,
$\epsilon$ the efficiency of forming stars out of this material, and
$M_{\ast}({\rm III})\sim 250 \,\msun$ the typical mass of a Pop~III
star.  We assume that roughly one-half of the mass that has been swept
up by the end of the energy-conserving (Sedov-Taylor) phase ends up in
the dense shell
\begin{equation}
    \Delta M_{\rm shell} \sim 0.5 \frac{4\pi}{3}R^3\rho_0
    \sim 0.5 \frac{E_0}{u_{\rm sh}^2}
     \;,
\label{eqn:shellmass}
\end{equation}
where $E_0\sim 10^{53}$erg is the kinetic energy of a PISN
\citep{FryWooHeg01}.  Taking the shock velocity at the onset of the
radiative phase to be $u_{\rm sh}\sim 300$~km~s$^{-1}$, we thus
estimate $\Delta M_{\rm shell} \sim 10^{4.5}\;\msun$.  In
equation~(\ref{eqn:efficiency}), we conservatively assume that only a
small fraction of the dense material in the shell is available to form
stars, $\epsilon \sim 0.01$, and we therefore predict the efficiency
of forming Pop~II.5 stars to be small: $\eta \sim 10^{2}\epsilon\sim 1$.

We would like to stress that the considerations presented here can
only serve to physically motivate the possibility of such an
intermediate stellar population. To fully explore the viability of our
proposal, we plan to investigate the propagation of a VMS SN with
numerical simulations (for numerical work on shock-induced star
formation in the present-day universe, see \citealt{VanCam98} and
references therein).

\subsection{Implications}

We propose that the three stellar populations discussed here
can form only in certain of the epochs introduced in \S~2. According
to our metallicity criterion, Pop~III star formation occurs exclusively
during the first two epochs. As we have argued above, the formation of
Pop~II.5 stars is induced by the death of a Pop~III star. Only the second
epoch, however, is conducive to such a triggered star formation mechanism.
In epoch 1, Pop~III stars form in low-mass halos that are already severely
disrupted by a single PISN. It is only later, in epoch 2 of the cosmic
star formation history, that the then substantially more massive halos can
survive a VMS explosion, and consequently enable the triggered formation
of intermediate-mass, Pop~II.5 stars. To see this more clearly, consider
the binding energy of a halo of mass $M$, collapsing at redshift $z$
\citep{BarLoe01}:
\begin{equation}
E_{b} \simeq 10^{54}{\rm erg} \left({M \over 10^{8}\;\msun}\right)^{5/3}
        \left({1+z \over 25}\right) .
\end{equation}
Using equation (1), we evaluate this expression
for $M \simeq M_{\rm crit}(z\simeq 25)$, resulting in
\begin{displaymath}
E_b \simeq \left\{
\begin{array}{ll}
10^{49}{\rm erg}   &  \mbox{for H$_{2}$ cooling}\\
10^{53}{\rm erg}   &  \mbox{for H cooling}\\
\end{array}
\right. \mbox{\ .}
\end{displaymath}
It is evident that the criterion for the formation of Pop~II.5 stars,
$E_b \ga E_0 \sim 10^{53}$erg, is fulfilled in halos that cool via
atomic hydrogen lines, but not in those that cool via H$_{2}$.

Within the framework presented in this paper, Pop~II stars can clearly
form in epoch 3, at which point we assume that most of the IGM is
enriched to a level in excess of $Z_{\rm crit}$.  It is likely,
however, that Pop~II star formation already ensues during epoch 2. The
metal enrichment due to VMSs is so efficient that the host system is
expected to reach the critical metallicity well before the general IGM
does. In fact, one can estimate that already {\it one single} VMS
could enrich the $\sim 10^7\,\msun$ of gas in the typical star-forming
halo in epoch 2 to the critical level:
\begin{eqnarray}
  {Z_{\rm halo}} &=& {M_Z \over M_B}
\nonumber \\
   &=& {\sim 10^2\;\msun \over \sim 10^7\;\msun}\sim 10^{-3.3}\zsun\ga Z_{\rm crit}
\nonumber \;,
\end{eqnarray}
where we have assumed for simplicity that none of the metals have escaped
the halo, corresponding to the limiting case of $f_{\rm mix}=0$
(see \S~2.3).

During epoch 2, therefore, we expect all three stellar populations to
occur almost simultaneously in a given star forming system. In Table
1, we summarize the association of the various stellar populations
with the distinct epochs of star formation.

We now tentatively suggest how these stellar populations might relate
to the different classes of SNe as proposed by \citet{QiaWas02}.  
These authors (henceforth QW) account
for the observed abundance patterns in low-metallicity Galactic
halo stars within the framework of three distinct classes of SNe. At
the beginning of the nucleosynthetic chain, a generation of very
massive stars explodes to enrich the primordial material mostly with
Fe, but producing no elements beyond the iron peak. It seems natural
to identify the QW VMS SNe with the Pop~III PISNe discussed here.
According to QW, at some later stage in the chemical enrichment
history, more normal, type II, SNe begin to contribute nucleosynthetic
products. More specifically, QW distinguish between a class of
high-frequency and low-frequency type II SNe, named SN~II({\it H}) and
SN~II({\it L}), respectively.  The {\it H} events produce mainly the
heavy {\it r}-process elements but no Fe, whereas the {\it L} events
contribute mainly the light {\it r}-process elements as well as Fe.
These two classes of SN~II are postulated on purely phenomenological
grounds, and it is presently not known what kind of progenitor stars
could physically give rise to them. There are, however, some
indications that the {\it H} events correspond to the explosion of
comparatively low-mass stars that are only barely able to reach the SN
stage (see QW and references therein).  Given that Pop~II.5 stars are
more massive than Pop~II stars by $\sim$one order of magnitude, we
speculate that Population~II.5 might predominantly give rise to
SNe~II({\it L}), whereas Population~II could lead to both the {\it H}
and {\it L} events. Again, we summarize these proposed identifications
in Table 1.

Regardless of the detailed correspondence discussed above, our general
model for the character of star formation at high redshift seems to
provide a promising astrophysical context for the QW proposal.  More
work, both observational and theoretical, is required to explore this
possible connection further.

\section{Observational Consequences}

\subsection{Supernova Rates as Star Formation Tracer} \label{sec:SNrate}
\subsubsection{Modeling}

The cosmic supernova rate as a function of redshift is potentially an
excellent tracer of the different modes of star formation envisaged in
this paper.  A number of authors have previously advocated this method
of tracing star formation, predicting supernova rates based on the
metallicity of the IGM at high redshifts \citep{MirRee97}, or on the
empirically determined star formation rate out to $z\sim 5$
\citep*[e.g.,][]{MadDelPan98,Sadetal98,DahFra99}.  In our study, we
use theoretical estimates of the cosmic star formation rate to predict
the number of supernovae from VMSs and from regular stars (Pop~II) out
to much higher redshifts.  Observational determinations of these rates
will constrain the three epochs of star formation which we are
proposing.  Including observational selection functions is beyond the
scope of this paper.  We will, however, present rough estimates of the
flux limit and wavelength coverage needed in order to observe SN
explosions at high redshift. The observational capability of NGST has
recently been studied for high-$z$ type~II SNe \citep{DahFra99}, and
in a preliminary fashion for PISNe from VMSs \citep{Hegetal02}.

We use the results of recent cosmological simulations by
\citet{SprHer02} to model the Pop~II star formation history of the
universe.  These simulations agree roughly with observations of the
star formation rate out to $z\sim5$, but extend to much higher
redshifts than are currently probed by direct observations, and also
reproduce the mean density of stars in the universe.  These authors
provide a simple analytical fit to the SFR as follows:
\begin{equation}
    \psi_*(z; {\rm II}) = K {b \exp [a (z-z_m)] \over
        b-a+a \exp [b (z-z_m)]} \;,
\end{equation}
where the normalization is $K=0.15 \;\msun {\rm yr^{-1}Mpc^{-3}}$ and
the fitting parameters are $a=3/5$, $b=14/15$, and $z_m=5.4$.  The
time delay between the formation and explosion of a massive star is
$<20 \;{\rm Myr}$, and therefore the Type~II supernova rate should
trace the star formation rate almost exactly.  For a Salpeter IMF,
there is one Type~II supernova (SN~II) per $\sim 150 \;\msun$ of star
formation, and we consequently divide the SFR by this factor (which we
call $M_{\rm SN}({\rm II})\;$) to obtain the Pop~II SN rate.
\citet{HerSpr02} find that the SFR at $z<z_m$ in their simulations
actually has a power-law dependence on the expansion rate, rather than
evolving exponentially with redshift; however the fit given in the
above equation is sufficient for our present purposes.

For the PISN rate we use the VMS star formation rate in
Figure~\ref{fig2}.  The typical lifetime of a VMS is $\sim 3 \;{\rm
Myr}$ so we can again assume the SN rate traces the SFR closely.  We
take typical VMSs to have masses of $250 \;\msun$ and assume they {\it
all} die as supernovae to calculate the VMS supernova rate (i.e.\
$M_{\rm SN}({\rm III})=250 \;\msun$).  VMSs are predicted to end their
lives as pair-instability supernovae if $M<260 \;\msun$, and these
should be very energetic and luminous events.  Bear in mind, however,
that the mass function, and hence end state of Pop~III stars is still
somewhat uncertain, making this supernova rate a crude estimate.

We also calculate the rate of Pop~II.5 supernovae, using the result
from the last section that the Pop~II.5 SFR is directly proportional
to the Pop~III SFR, with constant of proportionality $\eta \sim 1$.
Due to the uncertain nature of the Pop~II.5 IMF we must make simple
assumptions to get the mass in stars formed per supernova.  The
characteristic stellar mass was shown to be about $10\;\msun$, so we
assume that half of the stellar mass goes into stars that can explode.
We further assume that the average mass of a Pop~II.5 supernova
progenitor is $\sim 20\;\msun$.  From this we find that
$M_{\rm SN}({\rm II.5})\sim40 \;\msun$ for Pop~II.5 supernovae.  We
expect these supernovae to be of Type II, although with very low
metallicity.

We convert these rates per unit volume into rates observed per unit
time and per unit redshift interval:
\begin{eqnarray}
  {{\rm d}N \over {\rm d}t_{obs}dz} &=&
  {{\rm d}N \over (1+z){\rm d}t_{em}{\rm d}V}{{\rm d}V \over {\rm d}z}
\nonumber \\
   &=& {1 \over M_{\rm SN}}{\psi_*(z) \over (1+z)}{c\over H_0}
        {4\pi\ f_{sky} r^{2}(z) \over \sqrt{\Omega_m(1+z)^3+\Omega_{\Lambda}}}
\;.
\end{eqnarray}
Here ${\rm d}V$ is a comoving volume element, $M_{\rm SN}$ is the mass
in stars formed per supernova for the relevant stellar population,
$f_{sky}$ the fraction of the sky observed, and ${c/H_0}$ the Hubble
length.  Note that the Pop~II.5 and Pop~III supernova rates differ
only in their respective values for $M_{\ast}$ calculated above.  The
respective time intervals in the observer and emitter frames, ${\rm
d}t_{obs}$ and ${\rm d}t_{em}$, are related by the factor of $(1+z)$
in the denominator.  The comoving distance to redshift $z$ is
\begin{equation}
  r(z) = {c\over H_0}
         \int_0^z {{\rm d}z' \over \sqrt{\Omega_m(1+z')^3 +\Omega_{\Lambda}}}
  \;.
\end{equation}

\subsubsection{Results}

\begin{figure} 
\plotone{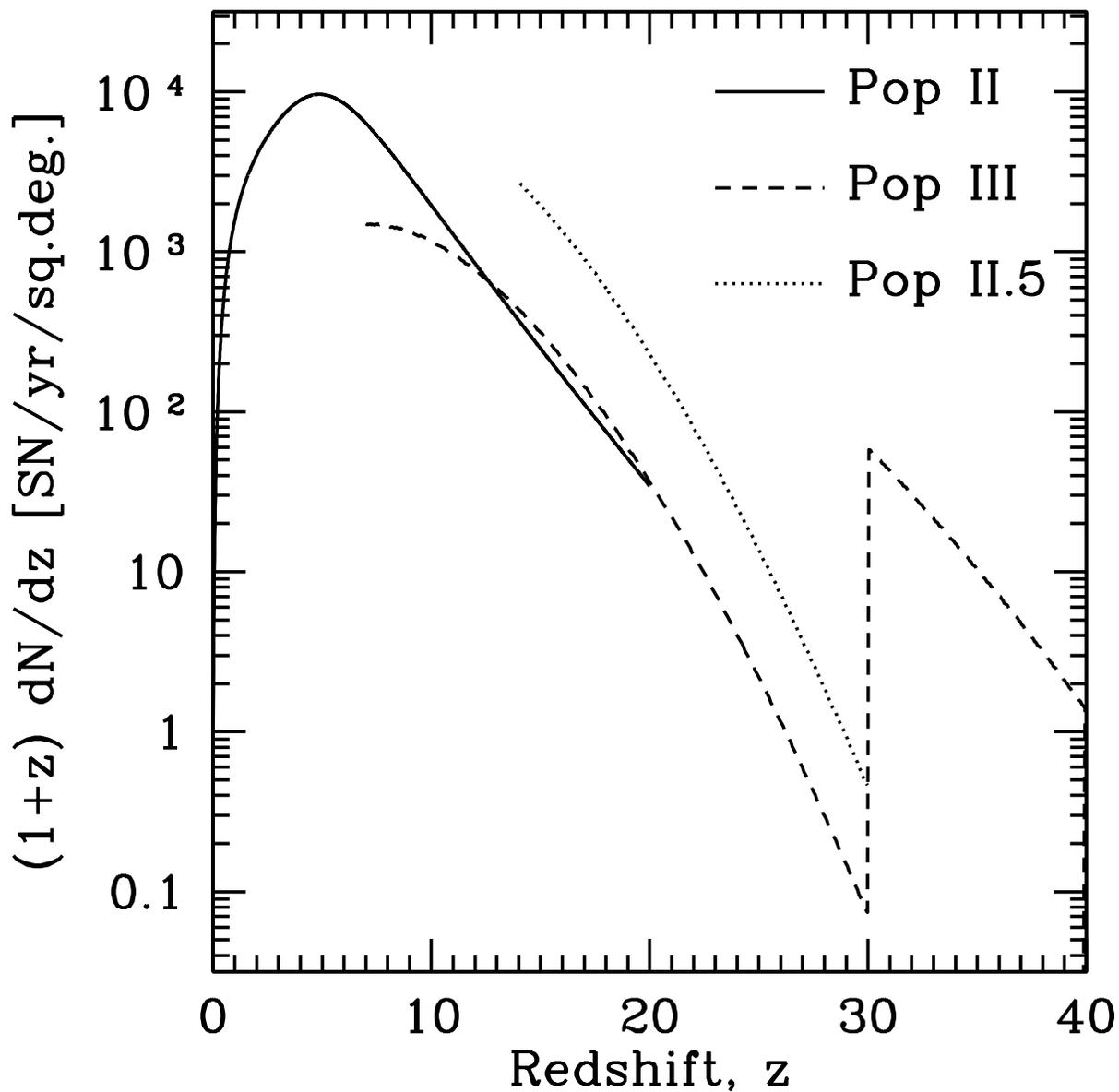}
\caption{Supernova rate as a function of redshift for
Type~II and VMS supernovae.  Pop~II and Pop~II.5 refer to Type II
supernovae from Pop~II and Pop~II.5 stars, respectively.  Pop~III
refers to VMS supernovae.  The assumed star formation rates
are detailed in the text.  We show the number of supernovae per
observed year in one square degree of the sky, plotted as
a function of redshift.
\label{fig:snr}}
\end{figure}

The results for our cosmological parameters are shown in
Figure~\ref{fig:snr}, where we plot the number of supernovae that
could be observed per year and per square degree.  Note that this
figure does not incorporate magnitude limits or observational
selection functions, but does simply show the predicted number of
supernovae that can in principle be observed.

The redshift distribution of SNe follows the star formation rate,
peaking at $z\sim5$ for Pop~II, and showing the same break at $z=30$
for PISNe as in Figure~\ref{fig2}.  We expect the transition from
predominantly PISNe to Pop~II SNe to occur around a redshift of $z\sim
15-20$ (see Figure~\ref{fig:metal}), and we therefore show the rates for
both types of supernova in this redshift range.  As this transition
occurs, we expect the Pop~II SN rate to gradually drop below the solid
line towards higher redshifts as Pop~II star formation switches off.
Similarly we expect the Pop~III (Pop~II.5) rate to drop below the
dashed (dotted) line towards lower redshifts as Pop~III star formation
ceases.  We do not expect any VMSs at $z\la 15$, but we extend
our calculation down to $z=7$ because the reionization of the universe
at roughly this redshift provides a robust lower limit for the end of
VMS formation \citep[e.g.,][]{Ohetal01}.  That the PISN and
Pop~II SN rates are so similar at the transition redshifts is because
the star formation rate for Pop~III at $z\sim15$ which we predict
analytically is similar to the rate (for Pop~II) found in the
simulations of \citeauthor{SprHer02} (see Figure~\ref{fig1}).  Again,
we emphasize that the break in the VMS supernova rate at $z=30$ is
artificially sharp, reflecting the abrupt transition in the underlying
star formation rate.

In terms of total numbers of supernovae, we predict $\sim 2\times10^6$
PISNe per year over the whole sky (above $z=15$), or about 50 per
square degree per year.  This still comprises only $\sim 0.4$\% of all
supernovae, as we find $\sim 4\times 10^8$ SNeII per year (all sky), a
value comparable to previous estimates \citep{MirRee97,MadDelPan98}.
This number can be understood by a simple order of magnitude estimate
as follows.  About 10\% of the baryons in the universe are in stars by
the present time, giving a local density in stars of
$\rho_* \simeq 0.1\Omega_b \rho_{\rm crit}
        \sim 6\times 10^8 \;\msun/{\rm Mpc^3}$ for our cosmological
parameters \citep{Pee93}.
The size of the observable universe is $\sim 10$ Gpc out to $z=5$
(most of the star formation happens later than $z=5$ because there is
very little physical time before this), so the total mass in stars in
this volume, ignoring redshift evolution effects, is
$\sim 10^{21.5}\;\msun$.
About 10\% of the mass in stars goes into stars which can explode, so
the mass in supernovae is $\sim 10^{20.5}\;\msun$.
The mass weighted average for a supernova progenitor's mass is
$\sim30\;\msun$, and the age of the universe is $14\;{\rm Gyr}$. The
number of supernovae per year is then
\begin{equation}
   \frac{{\rm d}N}{{\rm d}t_{\rm obs}}
   \sim {10^{20.5} \over 30\times14\times10^{9}{\rm yr} }
       \sim 7\times 10^{8} \ {\rm SNe\;yr^{-1}} \;,
\end{equation}
similar to our numerical integration of the calculated SN rate.  All
calculations of the SN rate based on the cosmic SFR are expected to
yield an answer around this value.

\subsubsection{Observational Prospects}
We now address the question of whether any of these supernovae at very
high redshift could be observed in the foreseeable future.  Type Ia
supernovae have a peak B-magnitude of $M_B \simeq -19$, with Type~II
supernovae being about 2.5 magnitudes fainter \citep[e.g.,][]{Fil97}.
VMSs are predicted to end their lives as pair instability supernovae
if their mass is in the range $140\;\msun < M<260\;\msun$
\citep{HegWoo02}, but they have not yet been observed and their
luminosity is not well known.  \citet{HegWoo02} have shown that
they can produce large masses of $^{56}$Ni during supernova
nucleosynthesis.  If the light curve is powered by $^{56}$Ni decay,
then these supernovae should be brighter than Type Ia's by a large
factor (the ratio of their $^{56}$Ni production).
Preliminary results from \citet{Hegetal02} show that reality is
somewhat more complicated, and these authors find that a PISN is only
a factor of a few brighter than a type Ia.  Because of this
uncertainty in luminosity, we will calculate the apparent brightness
of a type Ia SN at large redshifts, and scale this up for the brighter
PISN.  The apparent brightness in a flat universe is given by
\begin{equation}
   f_{\nu}(z) = { L_{(1+z)\nu} \over 4\pi (1+z) r^2(z)} \;,
\end{equation}
where $L_{\nu}$ is the monochromatic luminosity of the source at redshift
$z$.  For $z=20$, $M_B = -19$, and our cosmological parameters we
obtain
\begin{equation}
  f_{\nu}(\lambda_{\rm obs}\simeq9\mu {\rm m},z=20) \simeq 5\ {\rm nJy} \;.
\end{equation}
The observed wavelength of $9\mu {\rm m}$ is the redshifted $B$-band.  If
the flux does not drop off too much in the rest frame near-UV, then
there should be a similar flux level in the near-IR ($\lambda<5\mu {\rm m}$)
where NGST should have sub-nJy sensitivity \citep{BarLoe01}.  This indicates that even a Type Ia SN should
be observable with NGST out to $z\simeq 20$, and PISNe should be
easily observable if they are at least a few times brighter.  Type~II
supernovae, on the other hand, will be very hard to detect at these
high redshifts because they are typically a factor of 10 fainter.
For this reason, even though we predict Pop~II.5 supernovae to be 
more numerous than PISNe, we expect that they will not be nearly as
useful as observational probes of the high redshift universe.  Of 
course observing them would be useful for assessing the efficiency of 
low mass star formation, and for testing our shock compression model.

\citet{MarFer98} have discussed the feasibility of using gravitational
lensing magnification due to intervening mass concentrations to
observe high redshift supernovae more easily, and potentially bring
even type II supernovae within reach of NGST at $z\ga 10$.  Similar
considerations also apply to our calculations, and this may help in
obtaining detailed observations of PISNe.  It is important to be able
to identify them reliably, distinguish them from other types of
supernova, and measure their redshift if we are to use them for
cosmological purposes.  To be able to do this, we need a sample of SNe
that are bright enough for spectroscopic observations.  Since none
have been observed locally, we need the additional lensing
magnification to investigate SNe in detail at high redshift.

\subsection{Metal Poor Halo Stars}

If our mechanism for producing extremely low metallicity, low mass
stars (Pop~II.5 in our terminology) via shock compression occurs in
reality, then some of these stars should still be present in the halo
of our galaxy.  They would of course be very old ($\sim 14$\ Gyr), so
only the lowest mass stars with $M\la 0.8\;\msun$ would remain 
\citep*{Giretal00,SieLivLat02}.
We can use our model to predict how many such stars we would expect to
find in a typical Milky Way-sized halo and, given a model for the
stellar halo density profile, their number density in the solar
neighborhood (we use a similar approach to that of
\citealt{HerFer01}).  Significant uncertainties in this calculation
are the efficiency of producing Pop~II.5 stars (as discussed in
\S\ref{sec:pop2.5}), the fraction of them that end up in the halo as
opposed to the bulge, and their mass function.

The star formation rate of Pop~II.5 stars is given by $\psi_*(z; {\rm
II.5}) = \eta \psi_*(z; {\rm III})$ in the redshift range when they
form.  There is a slight difference to the way $\psi_*(z; {\rm III})$
is calculated for this problem, in that we are interested in the stars
that end up in the Milky Way (MW) halo.  Therefore, instead of the
usual Press-Schechter mass function in the \citet{SanBroKam02} model,
we use the Extended Press-Schechter mass function of halos at redshift
$z$ which will end up in a halo the mass of the MW at $z=0$
\citep{LacCol93}.

For a MW mass (total) of $M_{\rm MW}=10^{12}\;\msun$ we find that
$10^{7.5}\;\msun$ of Pop~III stars are formed in MW progenitor halos,
over the redshift range from $z=30$ to $z=15$ in which we expect
Pop~II.5 stars to form.  Interestingly, because the metal yield of
PISNe is $\sim 0.5$, this results in a MW metal enrichment of
\begin{equation}
   Z \sim 0.5\times \left({10^{7.5}\;\msun \over 10^{12}\;\msun}\right)
                      \left({\Omega_m \over \Omega_b}\right)
                      \left({\zsun \over 0.02}\right)
     \sim 10^{-2} \zsun\;,
\end{equation}
similar to the metallicity of the lowest metallicity globular clusters and
thick disk stars ([Fe/H]$=-2.2$, \citealt{FreBla02}).

For a given value of $\eta$, this mass in Pop~III stars gives us
the corresponding mass in Pop~II.5 stars:
\begin{equation}
  M_{\rm II.5} \simeq 10^{7.5} \;\msun \left({\eta \over 1.0}\right)
        \left({\eta_* \over 0.1}\right) \;.
\end{equation}
We now wish to convert this into the number density of Pop~II.5 stars
that could be observable in the MW halo today.  First, we have argued
that the characteristic mass of Pop~II.5 stars is $\sim10\;\msun$, so
most of these stars will have died long before the present.  We
parameterize this in a very simple way by assuming that some fraction,
$f_{\rm imf} \sim0.1$ of the mass in stars formed goes into stars with
$M<0.8\;\msun$.  A second poorly constrained process is where in the
MW these stars reside.  It is expected that most of the earliest
forming stars will be incorporated into the bulge of the MW, with a
small fraction in the halo \citep{WhiSpr00}.  We introduce $f_{\rm
halo} \sim 0.1$ as the fraction of Pop~II.5 stars which end up in the
halo.  Finally we make the reasonable assumption that most of the
Pop~II.5 stars remaining today are near the end of their lives (i.e.\
their mass function is increasing to higher masses at $M<1\;\msun$),
so we get the number of halo stars by dividing their total mass by
$0.8\;\msun$.  This gives us the total number of Pop~II.5 stars in the
MW halo:
\begin{equation}
  N_{\rm II.5} \sim 10^{5.5}\;\msun
        \left({\eta \over 1.0}\right)
        \left({f_{\rm halo} \over 0.1}\right)
        \left({f_{\rm imf} \over 0.1}\right)  \;.
\end{equation}
To relate this to the observed number density of low metallicity halo
stars we distribute the Pop~II.5 stars in a halo with density profile
$\rho(r)=\rho(r_{\odot})r_{\odot}^3/(r_c^3+r^3)$, where the solar
orbital radius, $r_{\odot}=8.0\;{\rm kpc}$ \citep{Rei93}, is assumed
much larger than the core radius, $r_c$, of the halo profile (needed
for finite halo mass).  This distribution only diverges
logarithmically at large radii, so the mass of the halo is quite
insensitive to the uncertain cutoff radius and core radius (we take
$r_{\rm cut}/r_{\rm core} = 300$ for definiteness).  Using this
profile, together with the total number of stars from the previous
paragraph, the number density of Pop~II.5 stars in the solar
neighborhood is
\begin{equation}
n_{\rm II.5} \sim 10 \;{\rm stars/kpc^3}  
        \left({\eta \over 1.0}\right)
        \left({f_{\rm halo} \over 0.1}\right)
        \left({f_{\rm imf} \over 0.1}\right)\;.
\end{equation}

We can make a crude comparison of this prediction with the results of
observational searches for metal poor halo stars
\citep[e.g.,][]{Beeetal00}.  Using their Figure 5, we see that most of
their stars are within 2 kpc of the sun, which we take to be the
radius of their survey (covering 1940 square degrees of the sky).
Also from that figure, they have $\sim 5$ stars with [Fe/H]$\leq
-3.5$, and none with [Fe/H]$\leq -4.0$.  If we identify these stars
with Pop~II.5, this gives them an observed space density of $n \sim 3
\;{\rm kpc}^{-3}$.  This is similar to our theoretical estimate above,
given the uncertainty associated with some of our parameters.

The observed stars all have a metallicity very close to our critical
metallicity, so it is not clear if they are very low metallicity
Pop~II stars or if they were formed by our Pop~II.5 mechanism.  We
have shown that, with reasonable modeling assumptions, it is possible
to produce these stars from primordial gas, and that their metallicity
could be due to pollution from Pop~III supernovae which triggered
their formation (it could also be partly due to pollution from swept
up gas as they travel through the galactic disk).  Current
observational searches are beginning to give us a clearer picture of
the number of very low metallicity stars in the galactic halo, and we
anticipate that this will continue with future observations.  On the
theoretical side, numerical simulations can in principle reduce the
uncertainties in our Pop~II.5 model by determining the values of
$\eta$, $f_{\rm halo}$, and possibly also of the IMF from the mass
spectrum of collapsing clumps, although this is a very challenging
problem to simulate accurately.

\section{Summary and Conclusions}

We have investigated the history of star formation in the high
redshift universe, focusing on the role played by very massive
Pop~III stars. We have argued that this history is shaped by
the various feedback effects exerted by those stars, resulting
in three distinct epochs of star formation.

The first impact Pop~III stars have on their surroundings is
radiative, as their soft UV flux dissociates the molecular hydrogen in
other minihalos \citep{HaiReeLoe97,HaiAbeRee00}.  Because of their
copious UV emission, and the small molecular fraction
($\sim10^{-4}-10^{-6}$) in minihalos, this photodissociation is likely
to happen quickly, and fairly completely.  A numerical investigation
of this process was performed by \citet{MacBryAbe01}, who simulated
this suppression of star formation in low mass halos by incorporating
a uniform soft UV background into a cosmological simulation.  While
this assumption does not allow them to follow the detailed build-up of
the UV radiation field as Pop~III star formation switches on, it does
give a good indication of the overall effect.  It was found that the
UV radiation can effectively suppress star formation due to molecular
cooling in low mass halos, delaying their collapse from redshift
$z\sim 30$ to $z\sim 20$.

It is encouraging that both analytic and numerical calculations agree
quite well.  In Figure~\ref{fig2}, we find that once a significant UV
radiation field has been set up by $z\sim30$, further star formation
is strongly inhibited until more massive halos start to form, and the
SFR does not recover to its $z\sim 30$ value until $z\la 20$.  This is
all in good agreement with the simulation results, except that
\citet{MacBryAbe01} did find that more massive halos of $M\ga
10^7\;\msun$ were able to retain some molecular hydrogen.  Hence,
these halos were still able to cool via molecular lines, indicating
that the SFR should not go all the way down to the atomic cooling
curve, but it is likely to drop most of the way.  Future simulations
will be able to follow this dissociation process in detail, taking
into account the effect of self-shielding, with a time-varying
radiation field generated by point sources at the sites of Pop~III
star formation. In summary, the radiative feedback from Pop~III stars
results in the transition, at $z\sim 30$, between the first two epochs
of star formation at high redshift.

The second impact Pop~III stars have on surrounding material is
chemical, and we identify this feedback as the main effect governing
the transition from forming predominantly very massive stars to
forming normal, lower mass ones
\citep[e.g.,][]{Omu00,NisTas00,Broetal01,Schetal02}.  With the
assumption that VMSs can only form in gas with a metallicity lower
than a critical value, $Z_{\rm crit}\sim 10^{-3.5}\zsun$, we estimate
that most of the star-forming material in the IGM has reached this
level in the redshift range $z\sim 15 - 20$, which delineates the
second transition in the star formation history, between epochs 2 and
3.

An important issue is whether such a well-defined transition redshift
between massive and more normal star formation actually occurs.  This
depends on how synchronized different parts of the universe are in
crossing the critical metallicity threshold, which in turn depends on
how well the supernova ejecta are mixed through the IGM at high
redshifts.  It has been suggested \citep{MadFerRee01} that enrichment
becomes more synchronized and uniform at higher redshifts.  Dark
matter halos are of much lower mass and have shallower potential
wells, making it easier for metals to escape the halo.  Supernovae
from VMSs are predicted to be very energetic, with explosion energies
significantly larger than the binding energy of halos at $z\ga 20$
\citep{BarLoe01}.  Furthermore, the universe is much denser at high
redshift and halos are consequently much closer together, again making
it easier to pollute the IGM uniformly with metals.  It is also
possible, however, that enrichment at high redshifts is similar to
that at low redshifts, where it has been shown to be very
inhomogeneous and incomplete
\citep[e.g.,][]{Gne98,CenOst99,Aguetal01a,Aguetal01b}.  If this were
the case, the transition redshift at which the mean mass-weighted
metallicity of the universe crosses the critical threshold would not
be a very meaningful quantity.  Which of these two cases is true is an
important and interesting question, and can only be addressed
theoretically by more detailed cosmological simulations which can
resolve star-forming halos at high redshifts.

The demise of Pop~III stars in SN explosions opens up the possibility
for a new stellar population which we have termed
Population~II.5. These stars are characterized by masses,
$M_{\ast}\sim 10\,\msun$, intermediate to that of Pop~III and Pop~II
stars, and by extremely low metallicities, below the critical value
that would normally lead to the formation of VMSs.  This possibility
relies on the gas in the vicinity of the SN being compressed in a
radiative shock, leading to a significant increase in density, and
thus possibly to a reduction of the fragment mass, $M_{\rm BE}\propto
\rho^{-1/2}$, to values much lower than the one realized in primordial
gas.  During an intermediate epoch, therefore, low- and high-mass
stars might form almost simultaneously.  
A similar idea had already been proposed in a prescient paper by
\citet{Cay86}, primarily to explain the so-called G-dwarf
problem, i.e., the observed lack of metal-poor stars in our Galaxy
\citep[see also][]{KasRee83,NakUme01}.

It is to be expected that the mass range of Pop~II.5 stars extends
down to $\la 1\;\msun$, and these stars at the low-mass end of the
Pop~II.5 IMF should still be around today.  Intriguingly and possibly
providing a case in point, the recent discovery of the extremely metal
poor star CS~29498-043 with [Fe/H]$=-3.7$ and a significant
overabundance of Mg and Si, has been tentatively interpreted as
hinting at a new class of stars \citep{Aoketal02}. Ongoing surveys of
extremely metal-poor halo stars, much improved in size and quality,
should soon be able to test our prediction further.

Many of the uncertainties in our argument have to be addressed with
detailed numerical simulations, and we plan to do so in future
work. The framework presented in this paper does provide a coherent
context for these numerical studies, and it highlights the important
physical questions that define the challenge of elucidating the end of
the cosmic ``dark ages''.

\acknowledgements{
We are grateful to Andrea Ferrara, Alex Heger, Marc Kamionkowski, Robert
Kirshner, Avi Loeb, Thomas Matheson, Mike Santos, Volker Springel, and 
Jerry Wasserburg for helpful discussions.
We thank the anonymous referee for comments that improved the
presentation of this paper.
This work was supported in part by NSF grants ACI
96-19019, AST 98-02568, AST 99-00877, and AST 00-71019.}

\end{document}